\documentclass[reprint,aps]{revtex4-2}
\usepackage{amsmath,mathtools}
\usepackage{amssymb}
\usepackage{amsthm}
\usepackage{graphicx}
\usepackage{physics}
\usepackage {xcolor}
\newcommand{\e}{\varepsilon}

\begin{document}
	
\title{Fermi Liquid near a q=0 Charge Quantum Critical Point}

\author{R. David Mayrhofer}
\affiliation{School of Physics and Astronomy and William I. Fine Theoretical Physics Institute, University of
Minnesota, Minneapolis, MN 55455, USA}

\author{Peter W\"olfle}
\affiliation{Institute for Theory of Condensed Matter and Institute for Quantum Materials and Technologies,
Karlsruhe Institute of Technology, 76131 Karlsruhe, Germany}

\author{Andrey V. Chubukov}
\affiliation{School of Physics and Astronomy and William I. Fine Theoretical Physics Institute, University of
Minnesota, Minneapolis, MN 55455, USA}

\date{\today}

\begin{abstract}
We analyze the quasiparticle interaction function (the fully dressed and antisymmetrized interaction between fermions) for a two-dimensional Fermi liquid at zero temperature close to a q=0 charge quantum critical point (QCP) in the $s-$wave channel (the one leading to phase separation). By the Ward identities, this vertex function must be related to quasiparticle residue $Z$, which can be obtained independently from the fermionic self-energy. We show that to satisfy these Ward identities, one needs to go beyond the standard diagrammatic formulation of Fermi-liquid theory and include series of additional contributions to the vertex function. These contributions are not present in a conventional Fermi liquid, but do emerge near a QCP, where the effective 4-fermion interaction is mediated by a soft dynamical boson.  We demonstrate explicitly that including these terms restores the Ward identity. Our analysis is built on previous studies of the vertex function near an  antiferromagnetic QCP [Phys. Rev. B 89, 045108 (2014)] and a d-wave charge-nematic  QCP [Phys. Rev. B 81, 045110 (2010)]. We show that for $s-$wave charge QCP the analysis is more straightforward and allows one to obtain the full quasiparticle interaction function (the Landau function) near a QCP.  We show that all partial components of this function (Landau parameters) diverge near a QCP, in the same way as the effective mass $m^*$,  except for the  $s$-wave charge component, which approaches $-1$. Consequently, the susceptibilities in all channels, except for the critical one, remain finite at a QCP, as they should.
\end{abstract}

\maketitle

\section{Introduction}
Fermi liquids -- systems of itinerant interacting fermions, form a class of materials that have been studied very thoroughly both theoretically and experimentally \cite{abrikosov1963methods,lifshitz1980statistical,pinesnozieres, baympethick}. It is well established  that the  observables  in a Fermi liquid have the same functional forms  as in a Fermi gas, but differ quantitatively. The theory of a Fermi liquid has been originally developed phenomenologically, based on conservation laws~\cite{Landau0,*Landau1,*Landau2}, and later reformulated microscopically, using the diagrammatic analysis and Ward identities~\cite{Pitaevskii1960,*Pitaevskii2011,lifshitz1980statistical,baympethick,Chubukov2018}.

In a microscopic description, the low-energy properties of itinerant fermions are determined by vertex function (the fully dressed antisymmetrized interaction) at strictly zero momentum and frequency transfer $\Gamma_{\alpha \beta, \gamma \delta}(K,P;K,P) = \Gamma_{\alpha \beta, \gamma \delta}(K,P)$,  where we use the abbreviation $K=(\bf k,\omega_k)$
\footnote{The vertex function, related to quasiparticle $Z$ by Ward identity, is often defined as $\Gamma^\omega$: the limit $Q \to 0$ of $\Gamma_{\alpha \beta, \gamma \delta}(K,P; k-Q, k+Q)$ with $Q = (q,\omega_q)$ taken at $q=0$ and infinitesimal $\omega_q$.  This definition implies that $\Gamma^\omega$ has no split low-energy poles in different  frequency half-planes.   One can easily verify that to satisfy this constraint there is no need to keep $\omega_q$ small but finite -- one can  just set both $\omega$ and $q$ to zero.}.
In a rotationally-invariant system in orbital and spin space, which we consider here, $\Gamma_{\alpha \beta, \gamma \delta}(K,P) = \Gamma_c  (K,P)\delta_{\alpha \beta} \delta_{\gamma \delta} + \Gamma_s (K,P) \sigma_{\alpha \beta} \sigma_{\gamma \delta}$. Partial components of $\Gamma_c$ and $\Gamma_s$ taken for fermions on the Fermi surface ($K = K_F, P = P_F$ where $K_F = ({\bf k}_F,0)$) determine the Landau parameters, which determine the renormalizations of thermodynamic properties, coming from fermions in the immediate vicinity of the Fermi surface.
For example, the Landau parameters are an essential ingredient to calculate the susceptibility of the system in the Fermi-liquid picture. 

Of particular interest to this work are the relations between $\Gamma_s$ and $\Gamma_c$  and the quasiparticle $Z$ factor -- the residue of the pole in the fermionic Green's function $G(k, \omega)$ at $k=k_F$. The $Z$ factor is not determined within Landau Fermi-liquid theory as it generally has contributions from fermions away from the Fermi surface.  However, within microscopic Fermi-liquid theory~~\cite{Pitaevskii1960,*Pitaevskii2011,kondratenko:1964,*kondratenko:1965} one can explicitly express
 $Z$ via either spin or charge component  of $\Gamma(K_F,P)$ with one of the momenta on the Fermi surface and one away from it. The relations are identical for $\Gamma_c$ and $\Gamma_s$ and are~\cite{abrikosov1963methods,Pitaevskii1960,*Pitaevskii2011,kondratenko:1964,*kondratenko:1965}
\begin{align}
\frac{1}{Z} &= 1 + 2 \int \Gamma_c(K_F,P) \{ G^2(P) \}_\omega \frac{d^3P}{(2\pi)^3}, \nonumber \\
\frac{1}{Z} &= 1 + 2 \int \Gamma_s(K_F,P) \{ G^2(P) \}_\omega \frac{d^3P}{(2\pi)^3}.
\label{eq:1}
\end{align}
where $G$  is the full Green's function. The quasiparticle residue $Z$ can also be obtained directly as the frequency derivative of the fermionic self-energy $\Sigma (k, \omega)$ at $k=k_F$ and $\omega \to 0$: $Z^{-1} = 1 + \partial \Sigma(k_F, \omega)/\partial \omega$ (we define self-energy via $G^{-1} (k, \omega) = \omega - \e_k + \Sigma (k, \omega)$). As long as the system is in the Fermi-liquid regime, $\Sigma (k_F, \omega) = \lambda \omega$ at the lowest frequencies, and $Z^{-1} =1+ \lambda$. The dimensionless $\lambda$ can be directly computed diagrammatically.  In this paper we obtain the vertex function $\Gamma_c (K,P)$ which satisfies Eq. (\ref{eq:1}).

This issue is most interesting for fermions near a quantum critical point (QCP), where $\lambda$ is large and Fermi liquid behavior holds only at the lowest frequencies. It is customary to describe the low-energy physics near a QCP within the effective model with the interaction $V_{eff} (K, P)$, mediated by soft fluctuations of the order parameter, which condenses on the other side of the QCP. This interaction involves low-energy fermions, hence both $K$ and $P$ can be placed near the Fermi surface. 

Systems at or near a QCP had been studied extensively in recent years~\cite{fitzpatrick2013,efetov2015,tsvelik2017,sung-sik2018,esterlis2021,shi2022,shi2024-2,guo2024},
with many recent studies focusing on the Ising nematic critical point ~\cite{metlitski2010, dalidovich2013, oganesyan2001, lawler2007,fradkin2010} as well as the transport properties of a system at a QCP ~\cite{lee2021,li2023,shi2024}.
In this work we consider a system near a momentum $q=0$ QCP in the charge sector -- an $s-$wave charge Pomeranchuk instability. This instability does not induce a homogeneous order parameter, bilinear in fermions, because the total electron density is conserved, but it leads to phase separation~\cite{dzyaloshinskii1974}. 

In a rotationally-invariant case, which we consider, $V_{eff} (K,P) = V_{eff} (Q)$, where $Q =K-P = ({\bf q}, \omega_q)$.
For a $q=0$ QCP, long-wavelength/low-frequency  bosonic fluctuations are relevant (small ${\bf q}$ and small $\omega_q$).
We follow earlier works (see e.g., Refs.~\cite{hertz1976quantum,metzner0,Nayak1994,avi}) and assume that the effective interaction can be approximated by a generalized Ornstein-Zernike form $V_{eff} (Q) = U(q)/(1 + U (q)  \Pi (Q))$, where $U(q)$ is a bare 4-fermion interaction and $\Pi (Q) >0$ is the polarization bubble at small momentum and frequency transfer. A charge instability develops when $U (0) <0$ at a point in parameter space where $U(0) \Pi (0) =-1$. Expanding around $Q=0$ and assuming $\omega_q \ll v_F q$, one obtains a Landau-overdamped effective interaction. In Matsubara frequencies, $V_{eff} (Q) \propto 1/(\xi^{-2} + (aq)^2 + \gamma |\omega_q|/q)$ where $a = O(1)$ and $\gamma \sim k^2_F/v_F$.
The Landau damping term comes from the expansion of $\Pi (Q)$ in frequency, and the $(aq)^2$ term comes from the expansion of both $U(q)$ and $\Pi (Q)$ in momentum.

The question we address here is whether near the charge QCP the vertex function $\Gamma (K,P)$ is the same as  $V_{eff} (K,P)$.  At a first glance, the answer is affirmative because the Ornstein-Zernike form of boson-mediated interaction can be explicitly derived by collecting series of renormalizations of the bare $U(q)$ by powers of $U(q)\Pi(Q)$.
We show below that the vertex function, obtained by antisymmetrizing the interaction and renormalizing the direct and antisymmetrized components  by series of $U(q) \Pi (Q)$, has the form
\begin{align}
\Gamma^{RPA}_{\alpha \beta, \gamma \delta}  = V_{eff} (Q)
\left(\delta_{\alpha \gamma} \delta_{\beta \delta} + \vec{\sigma}_{\alpha \gamma} \vec{\sigma}_{\beta \delta}
\right).
\label{ee_1}
\end{align}
We call this vertex function $\Gamma^{RPA}$, where RPA stands for random phase approximation, for consistency with notations in previous papers~\cite{Chubukov2018,maslov2010pomeranchuk,chubukov2014antiferromagnetic} and also because this vertex function can be explicitly obtained by summing up series of ladder and bubble diagrams made out of free fermions~\cite{dong1,*dong2}.
We see that the spin and charge components of $\Gamma^{RPA}$ are equal,
$\Gamma^{RPA}_c = \Gamma^{RPA}_s = V_{eff} (Q)$.  Hence if the first equation from (\ref{eq:1}) is satisfied, the second is also satisfied.

At second glance, there is an issue.  The same effective interaction $V_{eff}(Q)$ can be  used for the computation of the  fermionic self-energy $\Sigma (k,\omega)$.  Evaluating it, we obtain near a QCP $\Sigma (k, \omega) \approx \Sigma (k_F, \omega) =\lambda \omega$, where $\lambda$ is large and scales as $\xi$. This yields $1/Z \approx \lambda \propto \xi$.  Meanwhile, evaluating $1/Z$ from Eq. (\ref{eq:1}) using $\Gamma^{RPA}$ for $\Gamma$, we obtain (see Sec. IIIB) $1/Z \sim \sqrt{\lambda} \propto \sqrt{\xi}$.  The two forms of $Z$ clearly do not match.

We show in this work that the vertex function $\Gamma$ near a charge QCP differs from $\Gamma^{RPA}$ in two respects. First, the prefactor $a$ in $V_{eff} (Q)$ gets smaller when we include Fermi-liquid corrections.
This agrees with Refs.\cite{maslov2010pomeranchuk,wolfle2011,maslov2017gradient}.  We keep $a$ as a parameter, so this change does not crucially affect our analysis, although we will argue that the reduction of $a$ has an interesting consequence for the Fermi-liquid theory near a QCP. Second, we show that  $\Gamma$ differs from $\Gamma^{RPA}$ by an overall factor proportional to $\lambda$. Once we use full $\Gamma$ in Eq. (\ref{eq:1}), we recover the equivalence between $1/Z$ obtained from this equation and from the self-energy. We show that the  extra factor in $\Gamma$ comes from including additional ladder series of diagrams, which contain the convolutions of $ V_{eff} (Q)$ and two Green's function with exactly the same momenta and frequency, i.e., the terms $\int d^2q d \omega_q G^2(K+Q) V_{eff} (Q)$ (see Fig. \ref{ladder}  below). In an ordinary Fermi liquid away from a QCP, when the Landau damping term in $V_{eff}$ is irrelevant, the elements of these series vanish after frequency integration as the poles in the two Green's functions in the bubble with zero external momentum are in the same frequency half-plane. However, in a ``critical" Fermi liquid near a QCP, the Landau damping term in $V_{eff}$ becomes relevant.
Because this term has branch cuts in both  frequency half-planes, the frequency integral  $\int d \omega_q G^2(K+Q) V_{eff} (Q)$ does not vanish.  We show that near a QCP, each element of ladder series is $O(1)$, and their sum dresses $\Gamma^{RPA}$ by $O(\lambda)$. We obtain the explicit form of $\Gamma$, substitute it into Eq. (\ref{eq:1}) and reproduce $Z$ obtained from the self-energy.  We also touch on  Aslamazov-Larkin type diagrams for the vertex function and show that these diagrams are not relevant near a charge QCP.

Our consideration is an extension of earlier works~\cite{chubukov2014antiferromagnetic, chubukov2009ferromagnetic, maslov2010pomeranchuk}, in which the full $\Gamma$ was obtained in a critical Fermi liquid for a $d-$wave nematic order and $(\pi,\pi)$ antiferromagnetic order.  In these two cases, however,  the calculations and resulting expressions are quite involved, particularly for the $(\pi,\pi)$  case, where the full $\Gamma$ was obtained only in certain limits.
For the nematic case, the authors of ~\cite{chubukov2009ferromagnetic,maslov2010pomeranchuk} had to use approximate averaging procedures for higher order ladder diagrams, due to multiple $d-$wave form factors. For the $s-$wave  $q=0$ charge QCP, which we consider here, we obtain the full analytic form of $\Gamma$ and show explicitly that using this $\Gamma$ in Eq. (\ref{eq:1}) one obtains the same $Z$ as in the self-energy calculation. In this respect, the s-wave $q=0$  charge QCP better illustrates the essential physics at play than previously studied QCPs.

In addition, we also consider the susceptibilities $\chi_{n,c}$ and $\chi_{n,s}$ in spin and charge channels with different angular momentum channels, specified by $n$ and address the issue of how they evolve as the system approaches the $s-$wave QCP. Within Fermi-liquid theory~\cite{lifshitz1980statistical,baympethick,Chubukov2018}, $\chi_{n,c} \propto (m^*/m)/(1 +F_n)$ and  $\chi_{n,s} \propto (m^*/m)/(1 +G_n)$, where $m^*/m$ is the ratio of the effective and the bare fermionic masses, and $F_n$ and $G_n$ are Landau parameters -- partial components of $\Gamma_c (K_F, P_F)$ and $\Gamma_s (K_F,P_F)$, respectively (more accurate expressions are Eqs. (\ref{eq:x2}) below). Because critical behavior develops only in the $n=0$ ($s-$wave) channel, the susceptibilities $\chi_n$ must remain finite for all $n>0$, while $\chi_{n=0,c}$ must diverge. However, each $\chi_{n, c(s)}$ is proportional to $m^*/m$, which diverges near the QCP. To cancel this divergence, all Landau parameters other than $F_0$, must diverge at the same rate as $m^*/m$.

To confirm this, we consider a two-dimensional Fermi liquid with isotropic dispersion and
extract the Landau parameters from $\Gamma (K_F, P_F)$. We show that in a critical Fermi liquid, where $m^*/m$ is large, non$-s-$wave Landau parameters $F_n$ with $n >0$ and  $G_n$ with all $n$, diverge in the same way as $m^*/m$, rendering the susceptibilities $\chi_{n>0,s}$ and $\chi_{n,s}$ finite at an $s-$wave QCP. In addition, we  show that $1 + F_0$ vanishes at a QCP, but the slope is such that $1+F_0 \sim \xi^{-2} (m^*/m)$. As a consequence, the susceptibility $\chi_{n=0} \propto (m^*/m)/(1+F_0)$  in a critical Fermi liquid still scales as $\xi^2$, but the divergence is split between $m^*/m \sim \xi$ and $1/(1+ F_0) \sim \xi$. In the limit when the $(aq)^2$ term in $V_{eff}$ is nearly  eliminated by mass renormalization, one finds $m^*/m \sim \xi^2$, and $1+F_0$ remains finite down almost to  a QCP.  In this situation, the divergence of the $s-$wave charge susceptibility $\chi_{n=0}$ becomes entirely due to the divergence of $m^*/m$.

This feature, that all Landau parameters other than the one in the critical channel will diverge with the effective mass as the system approaches the QCP, is the defining feature of the critical Fermi liquid. This marks a clear departure of the conventional Fermi liquid regime, in which the Landau parameters are insensitive to the system's proximity to the QCP. 

The outline of the paper is as follows. In Sec. II, we review the properties of the vertex function in a conventional Fermi liquid, and present the vertex function in a Fermi liquid near a charge QCP within RPA  (we call it $V_{eff}$).
We then calculate both the quasiparticle residue $Z$ and effective mass in Sec. III. We do this in two ways:
first using fermionic self-energy (Sec. IIIA) and then using the Ward identity
with $V_{eff}$ for the vertex function  (Sec. IIIB). We show that the two results do not match when the system is close to the QCP.
In Sec IV, we consider the Fermi liquid renormalizations of $V_{eff}$ and the vertex function beyond RPA. We first show in Sec. IVA that $V_{eff}$ with Fermi liquid corrections included preserves essentially the same form as the RPA expression.
Then, in Sec. IVB, we identify the ladder series of diagrams of the vertex function, which are irrelevant in an ordinary Fermi liquid, but become relevant near the QCP, when the dynamics of $V_{eff}$ becomes essential.  We call this regime a critical Fermi liquid.
We calculate the contribution of these ladder diagrams explicitly and obtain the vertex function for the critical Fermi liquid. We demonstrate that the discrepancy in the calculation of $Z$ is resolved once one uses in the Ward identity the full vertex function instead of its RPA form.
In Sec. V, we obtain the Landau parameters as partial components of the  full vertex function, and use them to calculate the susceptibilities in channels with different angular momentum $n$.
We confirm that the $n=0$ charge susceptibility $\chi_{0,c}$ diverges as $\xi^2$, while all other $\chi_{n,c(s)}$ remain finite.
We present our conclusions in Sec. VI.

\section{Quasiparticle Vertex Function}

\subsection{Conventional Fermi liquid away from a QCP}
\begin{figure}[h]
	\begin{center}
		\includegraphics[scale=0.17]{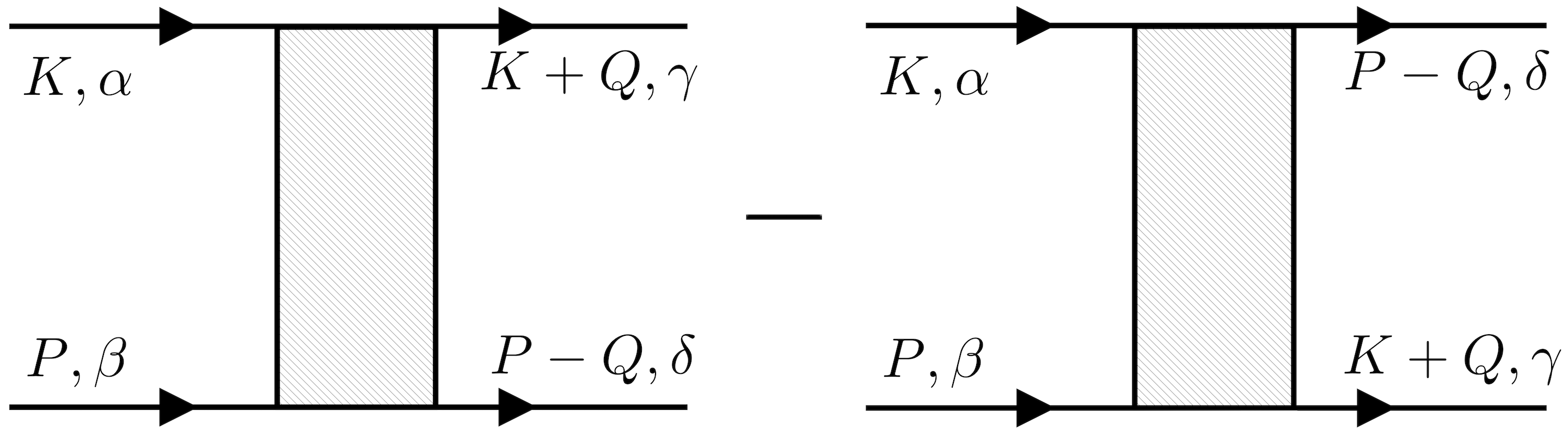}
		\caption{The total antisymmetrized vertex function.}
		\label{total_vertex}
	\end{center}
\end{figure}
We consider a 2D system of itinerant fermions with isotropic dispersion $\e_k$ and a static interaction $U(q)$, where $q$ is the momentum transfer between incoming and outgoing fermions with the same spin projection. By definition, the vertex function $\Gamma_{\alpha \beta,\gamma \delta}(K,P,Q)$ is a fully dressed, antisymmetrized 4-fermion interaction with vanishingly small momentum and frequency transfer $Q = (q, \omega_q)$.
At first order in interaction, we have (see Fig. 1)
\begin{align}
\Gamma_{\alpha \beta,\gamma \delta}(K,P,Q) =U(\vb{0}) \delta_{\alpha \gamma} \delta_{\beta \delta}  - U(\vb k - \vb p)
\delta_{\alpha \delta} \delta_{\beta \gamma},
\end{align}
where, since we consider vanishingly small momentum transfer, we have set $\vb q = 0$ in the interaction.
Using basic spin algebra, this function can be split into  spin and charge components in the particle-hole channel as
\begin{align}
\nonumber \Gamma_{\alpha \beta,\gamma \delta}(K,P,Q) &= \Gamma_c(K,P,Q) \delta_{\alpha \gamma} \delta_{\beta \delta} +\Gamma_s(K,P,Q) \vec{\sigma}_{\alpha \gamma} \cdot \vec{\sigma}_{\beta \delta}, \\
\nonumber \Gamma_c(K,P,Q) &= U(0) - \frac{1}{2} U(\vb k - \vb p),\\
\label{vertex} \Gamma_s(K,P,Q) &= - \frac{1}{2} U(\vb k - \vb p).
\end{align}

\begin{figure}[h]
	\begin{center}
		\includegraphics[scale=0.16]{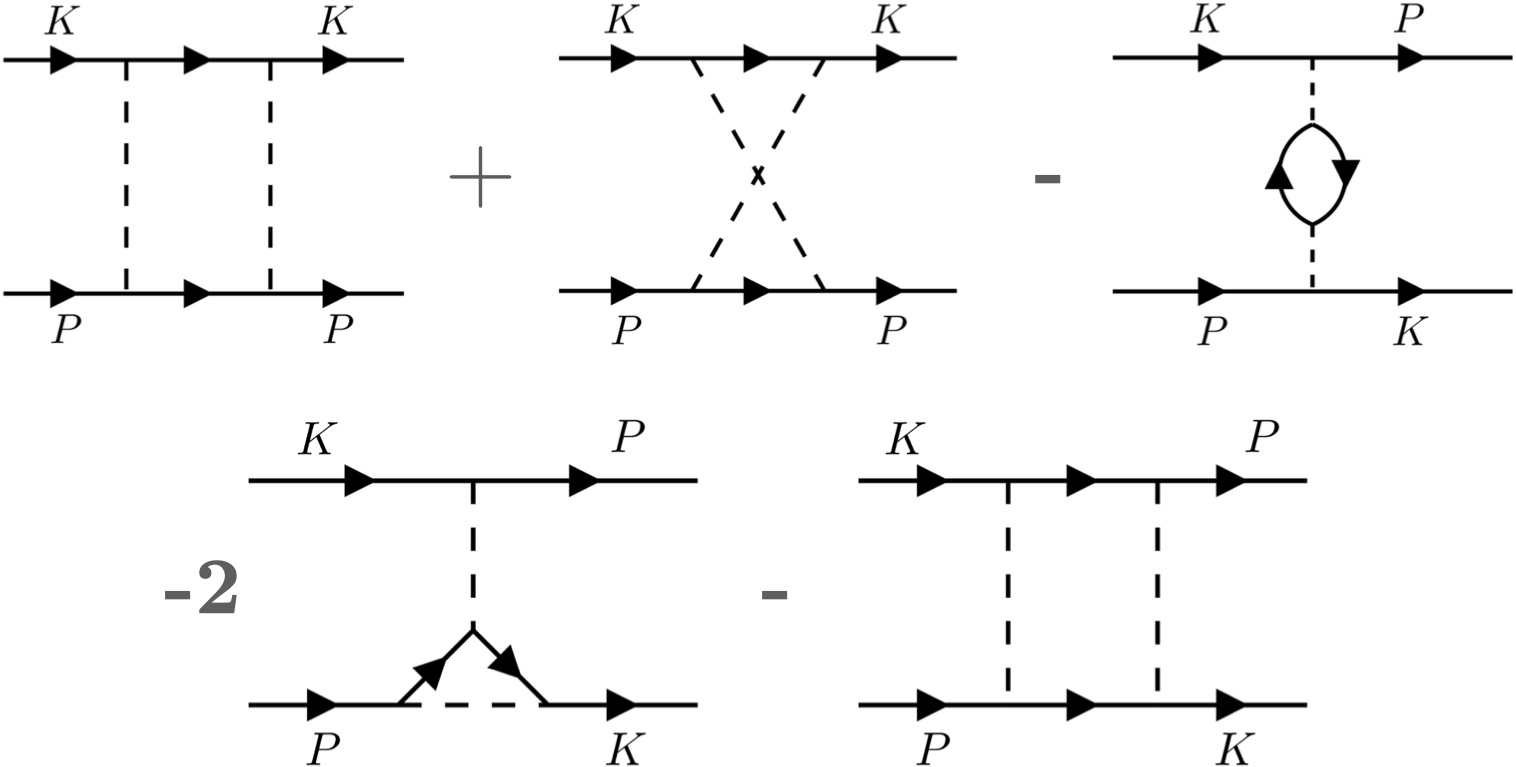}
		\caption{The diagrams that contribute to $\Gamma$ in the $\omega$ limit at second order in $U$.}
		\label{second_order}
	\end{center}
\end{figure}

We will ultimately need the vertex function in what is usually referred to as the $\omega$ limit, i.e. the limit when
$q = 0$ and $\omega_q \rightarrow 0$. The order of limits is relevant when considering higher order diagrams (Figs. 2 and 3), as there is a class of nominally singular diagrams that will not give a contribution in the $\omega$ limit. These ``forbidden'' diagrams contain a fermionic bubble with exactly zero momentum transfer: $\int d^2k d\omega G(k,\omega) G(k,\omega+\omega_q)$, which vanishes after the integration over frequency because the two poles are in the same frequency half-plane.  We note that the same holds if we set $\omega_q =0$, i.e., for this purpose the $\omega$ limit is equivalent to just setting $Q=0$ in all diagrams for the vertex function.

\begin{figure}[h]
	\begin{center}
		\includegraphics[scale=0.17]{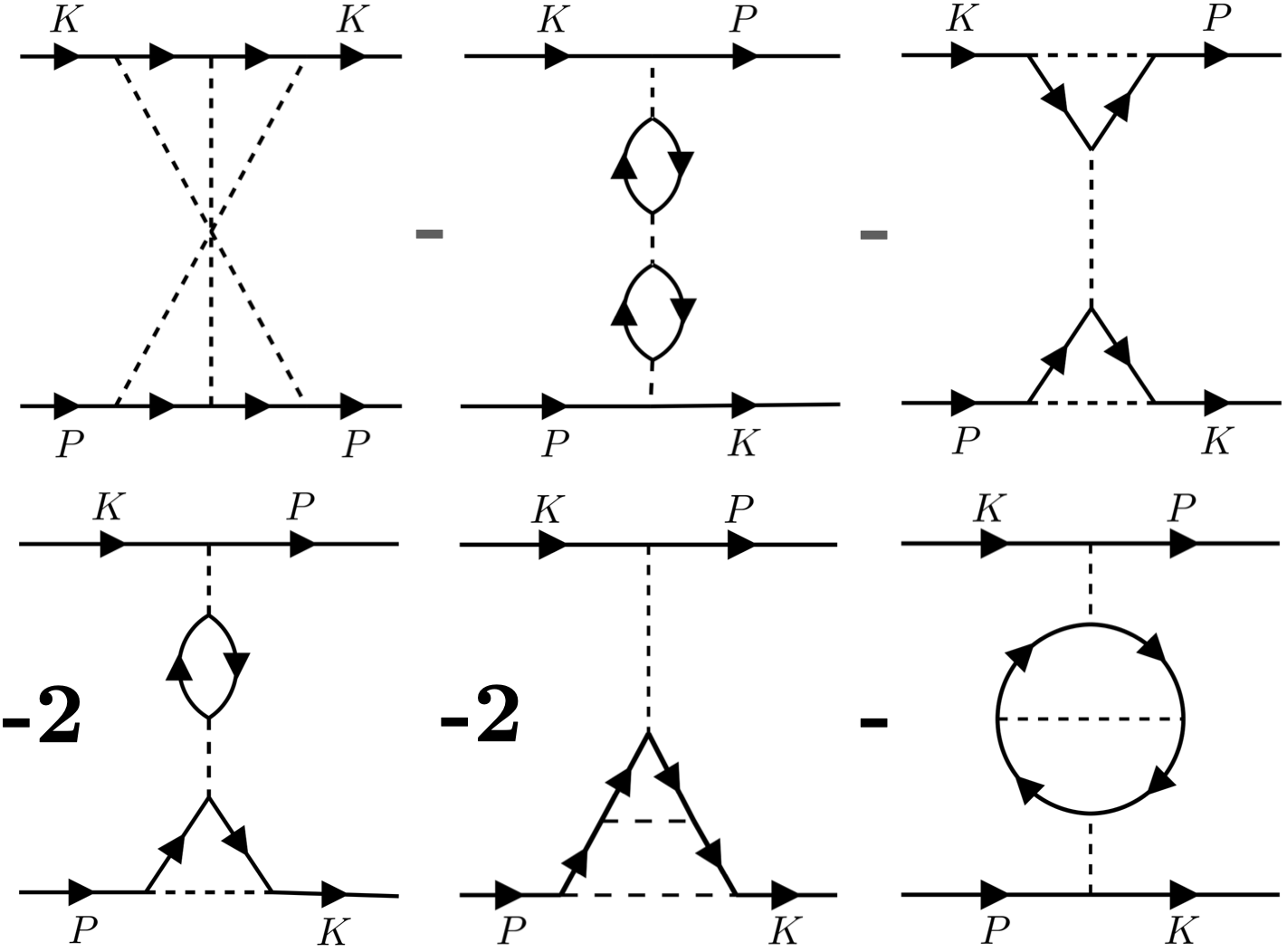}
		\caption{The diagrams that contribute to $\Gamma$ in the $\omega$ limit at third order in $U$. We neglect diagrams that contain one or more particle-particle bubbles.}
		\label{third_order}
	\end{center}
\end{figure}

Diagrams that contribute to the vertex function to order $U^2$ are shown in Fig. \ref{second_order}. The first and the last diagrams contain convolutions of the interaction and a particle-particle bubble $\Pi_{pp}(\vb k_F +\vb p_F)$, other diagrams contain convolutions of the interaction and a particle-hole bubble $\Pi_{ph}(\vb k_F - \vb p_F)$. The particle-particle bubbles are nonsingular near the q=0 particle-hole instability, and we neglect diagrams that contain them  for this reason.
This leaves three second-order diagrams that contain convolutions of the interactions and a $\Pi_{ph}(\vb k_F - \vb p_F)$.  Third-order diagrams of this type are shown in Fig. \ref{third_order}. We see that the series becomes repeating ladder and bubble diagrams.

For a generic $U(q)$, there is no easy way to sum up contributions to $\Gamma_{\alpha \beta,\gamma\delta} (K, P)$ to all orders in $U(q)$.  One can do this, however, if the interaction is sufficiently short-ranged such that $U(q)$ can be well approximated by a constant $U$.  Assuming that this is the case and using the summation procedure outlined in Refs. \cite{dong1,*dong2}, we obtain
\begin{align}
\nonumber \Gamma_{\alpha \beta, \gamma \delta} ^{RPA} (K_F,  P_F) =& \frac{U \vec{\sigma}_{\alpha \delta} \vec{\sigma}_{\beta \gamma} }{2\left( 1- U \Pi_{ph}(K_F - P_F)\right)} \\ 
&-\frac{U \delta_{\alpha \delta} \delta_{\beta \gamma} }{2\left( 1 + U \Pi_{ph}(K_F - P_F)\right)}.
\end{align}
where $K_F = ({\vb{k}}_F, 0)$. We define static $\Pi (q)$ as positive.
We call this vertex function $\Gamma^{RPA}$  because it is obtained by summing up series of ladder and bubble diagrams.

An s-wave instability in the charge channel occurs at negative  $U$, when the static $U\Pi(q \to 0) = -1$.
Near the instability the spin part of $\Gamma$ is non-singular and may be neglected. Keeping only the charge part, we obtain
\begin{align}
\label{chargerpa} \Gamma_{\alpha \beta, \gamma \delta} ^{RPA} \approx -
\frac{U \delta_{\alpha \delta} \delta_{\beta \gamma}}{2\left( 1 + U \Pi(K_F - P_F)\right)} = V_{eff}(K_F -P_F) \delta_{\alpha \delta} \delta_{\beta \gamma}.
\end{align}

\begin{figure}[t]
	\begin{center}
		\includegraphics[scale=0.03]{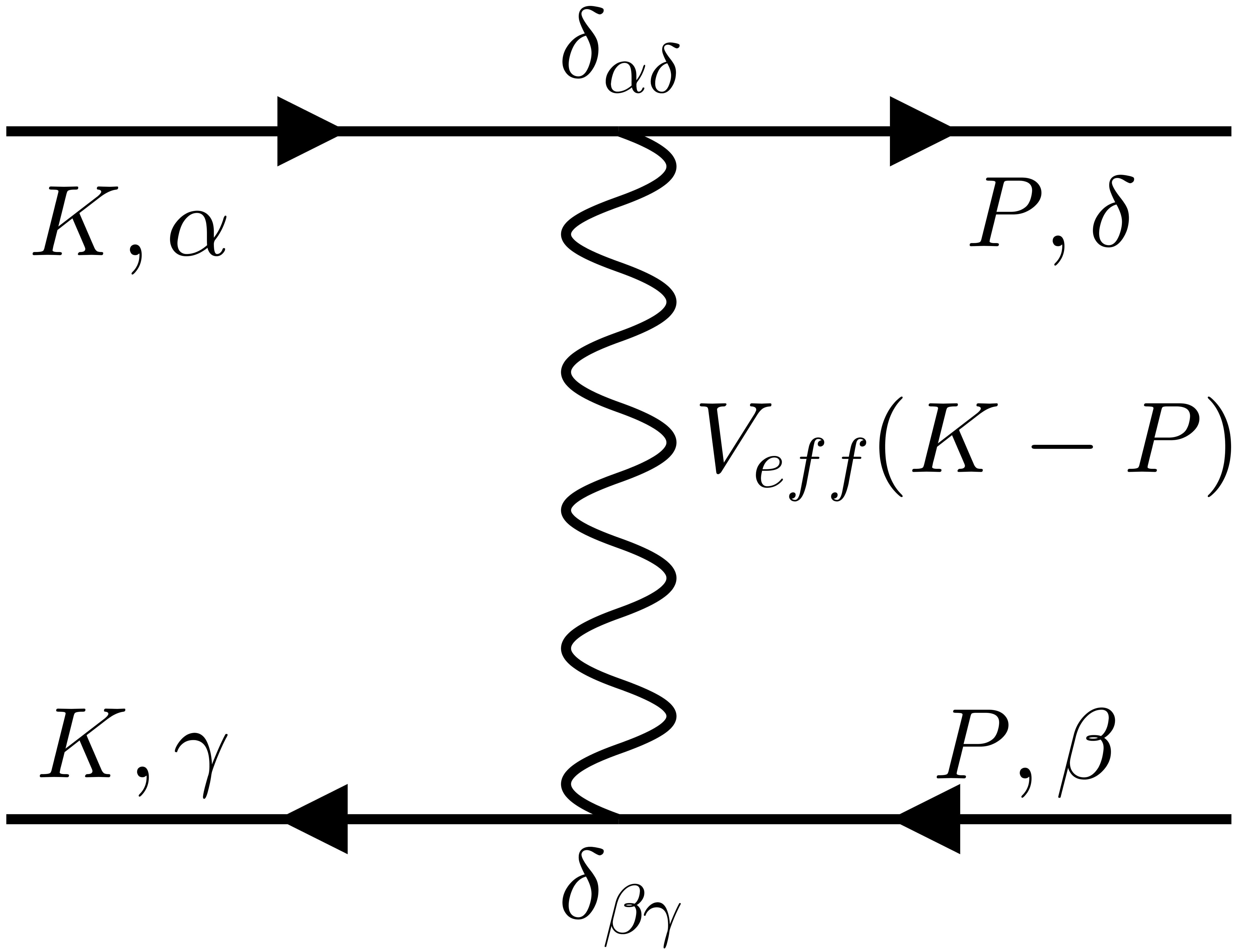}
		\caption{The effective interaction from Eq. (\ref{chargerpa}).}
		\label{rpa}
	\end{center}
\end{figure}
We emphasize that at the RPA level,  $\Pi_{ph}(Q)$ is constructed out of free fermions.  Neglecting spin component of $\Gamma^{RPA}$ is then legitimate only if there exists  a parameter range where charge component of $\Gamma$ is larger than its spin component, and at the same time Fermi liquid corrections are not strong enough to substantially affect $\Pi_{ph} (Q)$ compared to free fermion expression.  We discuss this below.

To obtain the functional form of $V_{eff} (K_F-P_F)$ at small but finite $\vb{q} = \vb{k}-\vb{p}$ and $\omega_q = \omega_k-\omega_p$, we assume that  relevant $v_F q$ are larger than relevant $\omega_q$. The dynamical part of $\Pi (Q)$ then has a form of a conventional Landau damping. Evaluating the polarization bubble for free fermions we then obtain
\begin{align}
\label{rpavertex}
 V_{eff}(Q) =\frac{k^2_F}{2N_F} \frac{1}{ \xi^{-2} + a^2 q^2 + \gamma \frac{|\omega_q|}{ |\vb q|}}
 \end{align}
where  $\gamma = k^2_F / v_F$, $a=O(1)$, and $\xi^{-2} = k^2_F(1-N_F |U|)$, where $N_F$ is the density of states at the   Fermi surface ($\int d^2 k/(2\pi)^2 = N_F \int d\e_k$).

Note that spin indices on the Kronecker deltas and the Pauli matrices in Eq. (\ref{chargerpa})  are different, compared to the first order expression, Eq. (\ref{vertex}). If we write this term out in terms of the same spin indices as Eq. (\ref{vertex}), we obtain
\begin{align}
\Gamma_{\alpha \beta, \gamma \delta} ^{RPA} = V_{eff} (K_F - P_F)
\left(\delta_{\alpha \gamma} \delta_{\beta \delta} + \vec{\sigma}_{\alpha \gamma} \vec{\sigma}_{\beta \delta}
\right).
\label{eq:6}
\end{align}
We see that the spin and charge components of the vertex function are equal.
This is consistent with the Fermi-liquid relations (\ref{eq:1}), which give
\begin{align}
\nonumber &\int \Gamma_c(K_F,P)\{G^2(P)\}_\omega \frac{d^3 P}{(2\pi)^3}  \\
= &\int \Gamma_s(K_F,P)\{G^2(P)\}_\omega \frac{d^3
P}{(2\pi)^3}.
\label{eq:5}
\end{align}
In a Fermi liquid far away from a QCP, this would represent a complex integral relation between $\Gamma_c(K_F,P)$ and $\Gamma_s (K_F,P)$, where $P$ is away from the Fermi surface. However, near a QCP, the dominant contribution to each integral comes from $P \approx P_F$ (more on this below).  In this case,  Eq. (\ref{eq:5}) requires $\Gamma_c(K_F,P_F)$ and $\Gamma_s(K_F,P_F)$ to be equal. This agrees with Eq. (\ref{eq:6}).

\begin{figure}[t]
	\begin{center}
		\includegraphics[scale=0.15]{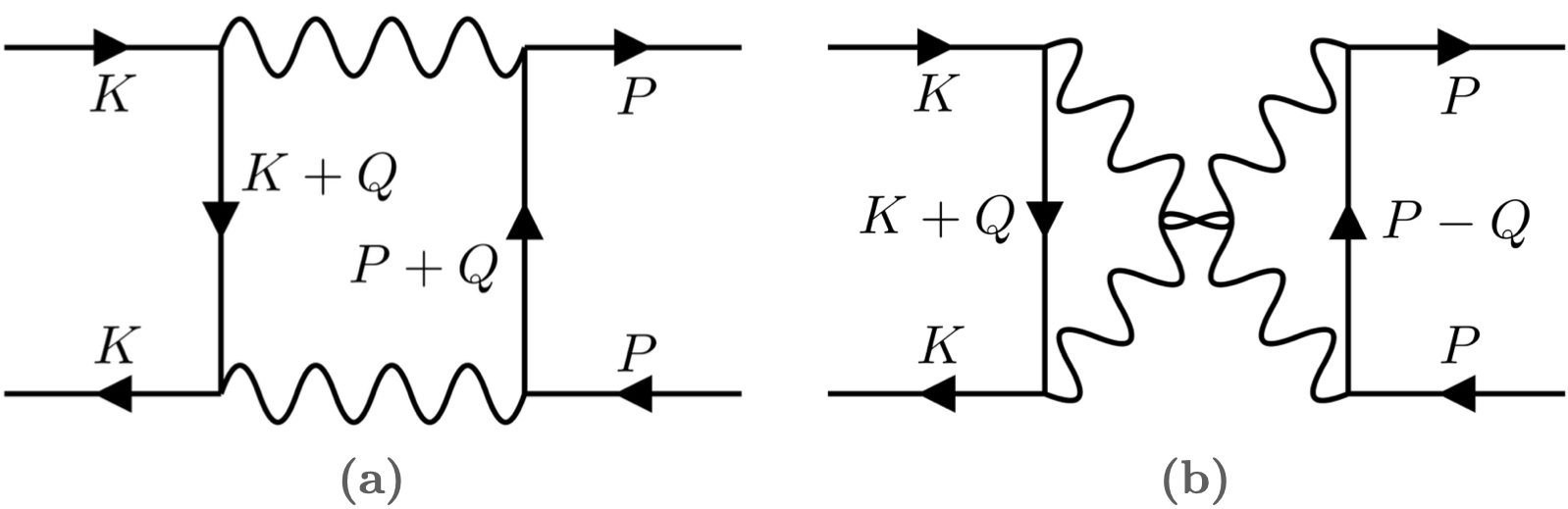}
		\caption{The two Aslamazov-Larkin type diagrams. The two cancel near a charge instability.}
		\label{aslamazov_larkin}
	\end{center}
\end{figure}

The equivalence between spin and charge component of $\Gamma_{\alpha \beta, \gamma \delta}^{RPA}$ does not hold for a $q=0$ spin QCP (the one towards ferromagnetism). In this case, spin and charge components of $\Gamma^{RPA} (K_F, P_F)$ differ by a factor of 3~\cite{chubukov2009ferromagnetic}.  The equivalence is restored once one adds Aslamazov-Larkin diagrams, Fig. \ref{aslamazov_larkin} (Refs.
\cite{chubukov2014antiferromagnetic,chubukov2009ferromagnetic,*betouras2014}).  For our case of a  charge QCP,  the two Aslamazov-Larkin contributions to the vertex function cancel each other~\cite{maslov2010pomeranchuk}. Indeed, the contributions from the two diagrams in Fig. \ref{aslamazov_larkin}  are
\begin{align}
\nonumber I_a =& \sum_{s,t} \delta_{\alpha s} \delta_{t \delta} \delta_{s \gamma} \delta_{\beta t} \\
&\times \int \frac{d^2q}{(2\pi)^2} \frac{ d\Omega}{2\pi} V_{eff}(Q)^2 G(K+Q) G(P+Q) \\
\nonumber I_b =& \sum_{s,t} \delta_{\alpha s} \delta_{t \beta} \delta_{s \gamma} \delta_{t \delta}\\
&\times \int \frac{d^2q}{(2\pi)^2} \frac{ d\Omega}{2\pi} V_{eff}(Q)^2 G(K+Q)G(P-Q),
\end{align}
For relevant $P$ on the Fermi surface we have
\begin{align}
G(P-Q) = \frac{1}{- i \omega_q + \vb{v}_F \cdot \vb q} = -G(P+Q),
\end{align}
One can then immediately verify that the contributions from the two diagrams are equal in magnitude but
opposite in sign, i.e., the sum of the two contributions vanishes. For the spin case,
where we have Pauli matrices at the vertices instead of Kronecker deltas, the two diagrams no longer have the same spin dependence, and the summation of the two yields a finite contribution.

\section{Quasiparticle residue $Z$ and effective mass $m^*$ within RPA}
\begin{figure}[h]
	\begin{center}
		\includegraphics[scale=0.025]{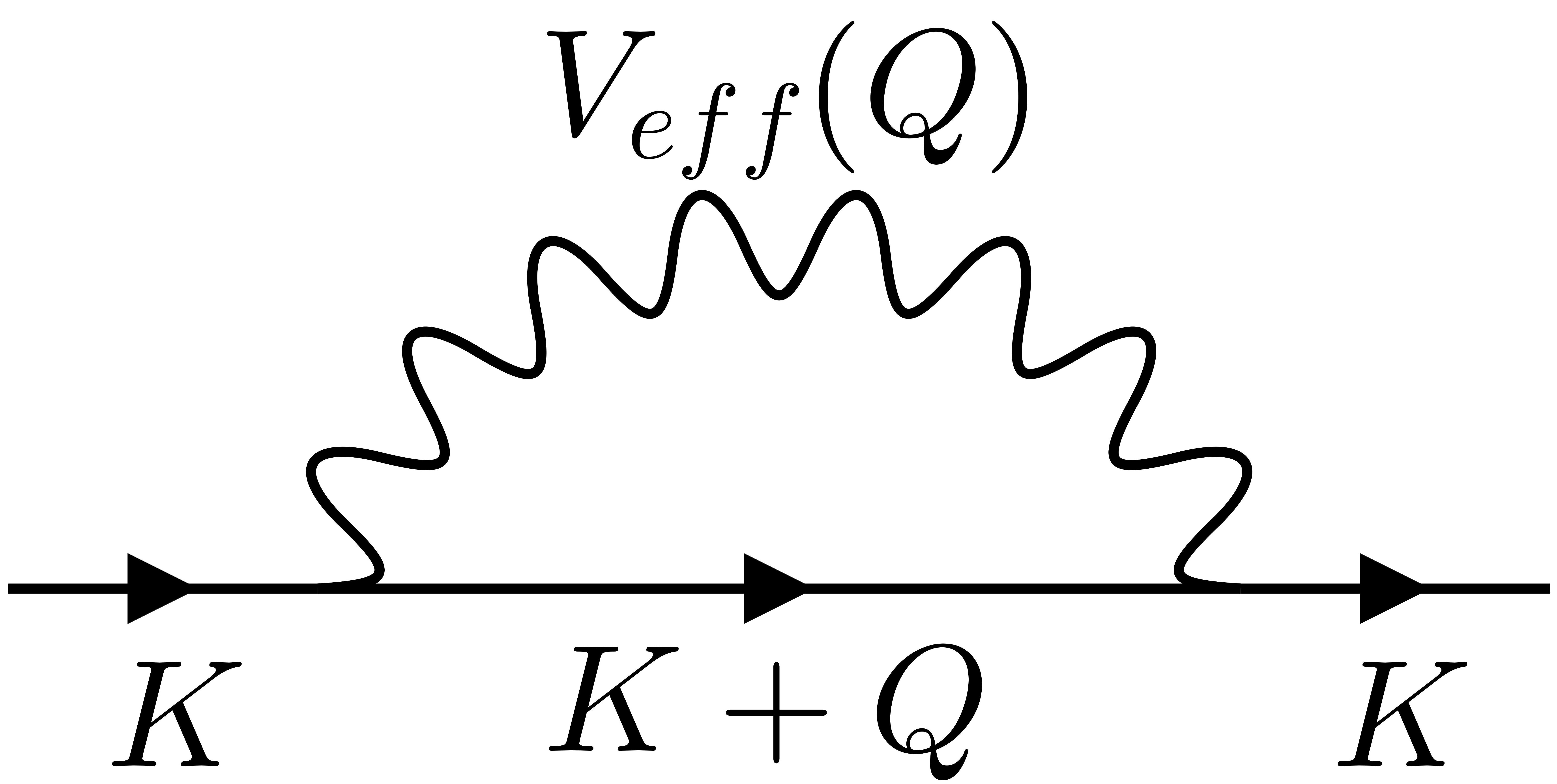}
		\caption{The one loop expression for the fermionic self-energy, with the RPA interaction $V_{eff} (Q)$}
		\label{self_energy}
	\end{center}
\end{figure}

For a generic Fermi liquid, fermionic propagator at $T=0$ at small $\omega_m$ and $k \approx k_F$ is
\begin{align}
G(K) = \frac{Z}{i \omega_k - k_F (k-k_F)/m^*}
\label{eq:g}
\end{align}
In this section we compute $Z$ and $m^*/m$ using Fermi liquid form of $G(K)$ and RPA form of the effective interaction.
We do computations in two ways: by evaluating fermionic self-energy and by using Eq. (\ref{eq:1}). We show that the results  do not match. In the next section we argue that to restore the equivalence one needs to evaluate the vertex function $\Gamma$ beyond RPA.

\subsection{$Z$ and $m^*/m$ from the self-energy}

The effective interaction $V_{eff} (Q)$ can be used to compute the fermionic self-energy. At one-loop order,
\begin{align}
\Sigma(k,\omega_m) = -\int \frac{d^2q}{(2 \pi)^2} \frac{ d \omega_q}{2\pi} V_{eff}(Q) G(K+Q)
\end{align}
Subtracting $\Sigma (k_F,0)$, which is irrelevant to our purposes, expressing $\Sigma(k,\omega_m)$ as $\Sigma(k,\omega_m)= \Sigma (k_F,0) + \delta \Sigma(k,\omega_m)$, we obtain for $\delta \Sigma(k,\omega_m)$:
\begin{widetext}
\begin{align}
\label{eq:8}
\delta \Sigma(k,\omega_m) &= -\int \frac{d^2q}{(2 \pi)^2} \frac{ d \omega_q}{2\pi} V_{eff}(Q) \left(G(K+Q)-G(K_F+Q)\right)\\
\nonumber &= - \frac{Z k^2_F }{2N_F} \int \frac{d^2q}{(2 \pi)^2} \frac{ d \omega_q}{2\pi} \frac{1}{\xi^{-2} + a^2 q^2 +
\gamma \frac{|\omega_q|}{q}} \frac{\e_k^* - i \omega_m}{\left(i(\omega_m + \omega_q) - \e_k^*
-\vb{v}_F^* \cdot  \vb{q} \right) \left( i \omega_q -\vb{v}_F^* \cdot  \vb{q} \right)}.
\end{align}

Let's set $\omega_m>0$ for definiteness.
The momentum and frequency integral in (\ref{eq:8}) is ultraviolet convergent and can be evaluated in any order. It is convenient to do momentum integration first.  There are two contributions to the momentum integral: one comes from the poles in the fermionic propagators, and the other from the pole in the effective interaction. We label them as $\Sigma_f$ and $\Sigma_b$. The two poles in the fermionic propagators are in different half-planes when $\omega_q$ is within the range  $-\omega_m < \omega_q < 0$. Let's direct $\vb{v}_F^*$ along $x$, i.e., write $\vb{v}_F^* \cdot \vb q = v_F^* q_x$. Typical $\omega_q$ in the pole contribution are of order $\omega_m$ and typical $q_x \sim \omega_m/v_F^*$. In the bosonic propagator, typical $q_y \sim \xi^{-1}$ and typical $\gamma |\omega_q|/|q| \sim k^2_F |\omega_m|/(v_F \xi^{-1})$.
At small enough $\omega_m <v_F \xi^{-1}/(k_F \xi)^2 < v_F \xi^{-1}$ we can set $\omega_q =0$ and $q_x =0$ in the bosonic propagator. Evaluating the pole contribution and integrating over $q_y$ in the bosonic propagator, we then find
 \begin{align}
\label{fermion_pole} \delta \Sigma_{f}(k,\omega_m) &= \frac{Z k^2_F }{2N_F v_F^*} i \omega_m \int \frac{dq_y}{(2\pi)^2}
\frac{1}{\xi^{-2} + (aq_y)^2} = i \omega_m \lambda,
\end{align}
where $\lambda = (\xi k_F/4a) (Z m^*/m)$

We next  calculate the contribution $\delta \Sigma_b$ from the pole in the effective interaction. Because the numerator in (\ref{eq:8}) already contains $\e_k^* - i \omega_m$, we set
$\omega_m$ and $ \e_k^*$ to 0 in the fermionic propagator. We then have
\begin{align}
\label{effective} \delta \Sigma_{b}(k,\omega_m) &= \frac{Z k^2_F}{2N_F}   (i \omega_m - \e_k^* )\int \frac{d^2q}{(2 \pi)^2}
\frac{ d \omega_q}{2\pi} \frac{1}{\xi^{-2} + (aq)^2 + \gamma \frac{|\omega_q|}{q}}
\frac{1}{\left( i \omega_q -\vb{v}_F^* \cdot  \vb{q} \right)^2}
\end{align}
Because the effective interaction does not depend on the angle between $\vb{v}_F$ and $\vb q$, we can do the angular integral first
\begin{align}
\int \frac{d \theta}{2\pi} \frac{1}{\left( i \omega_q - v_F^* q \cos \theta \right)^2} = -\frac{|\omega_q|}{\left(\omega^2_q + (v_F^* q)^2 \right)^{3/2}}.
\end{align}
\footnote{The angular integral also contains the term proportional to $\delta(\omega_q)$, but we can ignore it as we
 are considering the contribution from $\omega_q \gg \omega_m >0$.}
 Substituting this back into (\ref{effective}), we find
\begin{align}
\delta \Sigma_{b}(k,\omega_m) = -\frac{Z k^2_F}{2N_F} (i \omega_m - \e_k^* )\int_{0}^{\infty}
\frac{dq}{2\pi}\int_{-\infty}^{\infty} \frac{d\omega_q}{2\pi} \frac{q}{\xi^{-2} + (aq)^2 +
\gamma \frac{|\omega_q|}{ q}} \frac{|\omega_q|}{\left(\omega^2_q + (v_F^* q)^2 \right)^{3/2}}
\end{align}
Restricting integration over $\omega_q$ to positive frequencies and introducing polar coordinates
 $q = r \cos \phi$ and $\omega_q = v_F^* r \sin \phi$, we get
\begin{align}
\delta \Sigma_{b}(k,\omega_m) = -\frac{Z k^2_F}{v_F^* N_F } \left( i \omega_m - \e_k^* \right) \int \frac{dr d \phi}{(2\pi)^2}
\frac{\cos \phi \sin \phi}{\xi^{-2} +  (ar)^2 \cos^2 \phi + v_F^* \gamma \tan \phi}
\label{ee}
\end{align}
Integrating over $r$ and then over $\phi$, we obtain
\begin{align}
\label{eq:7}\delta \Sigma_{b}(k,\omega) &= Z \left(\frac{m^*}{m}\right)^{3/2}\left( i \omega_m - \e_k^* \right) f(\xi), \\
f(\xi)
&= -\frac{1}{4a}  \left [\frac{\Gamma^2(3/4)}{\sqrt{\pi}} - \frac{2 \Gamma^2(5/4)}{\sqrt{\pi}} \frac{m^*}{m(k_F\xi)^{2}}+
\mathcal{O}\left(\left(\frac{m^*}{m(k_F \xi)^2}\right)^2\right) \right],
\label{eq:7a}
\end{align}
\end{widetext}

where the $\Gamma$ above is the gamma function.
Combining $\delta \Sigma_f (k, \omega_m)$ and $\delta \Sigma_b (k, \omega_m)$ into $\delta \Sigma (k, \omega_m)$ and extracting $m^*/m$ and $Z$ from $G^{-1} (k, \omega_m) = i\omega_m -\e_k + \delta \Sigma (k, \omega_m) = (i\omega_m - \e^*_k)/Z$, we obtain two self-consistent equations for $Z$ and $m^*/m$. They are
 \begin{eqnarray}
\frac{1}{Z} &=& 1+\lambda  + f (\xi) Z \left(\frac{m^*}{m}\right)^{3/2} \nonumber \\
\frac{m^*}{m} &=& \frac{1}{Z} - f (\xi) Z \left(\frac{m^*}{m}\right)^{3/2}
\label{eq:s1}
\end{eqnarray}
 Solving this set together with Eq. (\ref{eq:7a}) we obtain
 \begin{equation}
 m^*/m = 1 + \lambda, ~ \frac{1}{Z} \approx  1 + \lambda + f(\xi) \sqrt{\lambda},~  f (\xi) \approx
  -\frac{1}{4a} \frac{\Gamma^2 (3/4)}{\sqrt{\pi}}.
  \label{eq:s1a}
  \end{equation}
To leading order in $\lambda$, we then have $m^*/m \approx 1/Z$.   We see that mass renormalization is entirely determined by the quasiparticle residue. This is the consequence of the fact that $\e_k$ itself changes by $\e_k (1 + f(\xi) Z (m^*/m)^{1/2})$. The correction term scales as $1/\sqrt{\lambda}$ and is small.

A closer look shows that we must be more careful when evaluating $\delta \Sigma_b (k, \omega_m)$.
In the integral over $r$ in (\ref{ee}), the dominant contributions come from $r\sim k_F \sqrt{m/m^*}$, and therefore, relevant $\omega_q \sim E_F \left(m/m^*\right)^{3/2}$. The Fermi liquid fermionic Green's function, given in Eq. (\ref{eq:g}), is only valid in the regime where the fermionic self-energy is approximately linear in frequency. This holds when frequency is smaller than $\omega_{FL}=(E_F/a)/(k_F\xi)^{3}$. Near the QCP, $m^*/m$ is large and relevant $\omega_q$ are above $\omega_{FL}$. At such frequencies we have to use quantum-critical, non-Fermi liquid form  $\Sigma(k=k_F,\omega) = i \omega_0^{1/3}|\omega_m|^{2/3}$ with $\omega_0 = \pi^3 E_F/(12 \sqrt{3} a^4)$ (see, e.g., ~\cite{avi} and references therein).
Substituting this form into the expression for $\Sigma_b$, we obtain instead of (\ref{effective})
\begin{align}
\nonumber \delta \Sigma_b (k,0) =& - \frac{k^2_F \e_k}{2N_F} \int \frac{d^2q}{(2 \pi)^2} \frac{ d \omega_q}{2\pi} \frac{1}{a^2 q^2 +
\gamma \frac{|\omega_q|}{q}} \\
&\times \frac{1}{\left(i |\omega_q|^{2/3} \omega^{1/3}_0
-\vb{v}_F \cdot  \vb{q} \right)^2},
\end{align}
Restricting the $\omega_q$ integral to positive values, we may replace the lower limit by $0$  if typical values of  $\omega_q$ are much larger than $\omega_{FL}$.  In our case $\omega_{q,typ}/\omega_{FL} \approx (48 \sqrt{3}/\pi^3) a^3 \left(k_F \xi \right)^3$, which we assume to be larger than one. Again evaluating the integral over the angle between $\vb v_F$ and $\vb q$ first, we now obtain
\begin{align}
\nonumber \delta \Sigma_b (k,0) =  \frac{k^2_F \e_k}{2N_F} \int \frac{dq}{2 \pi}& \frac{ d \omega_q}{2\pi} \frac{q}{a^2 q^2 + \gamma \frac{|\omega_q|}{q}}\\
\times & \frac{\omega^{2/3}_q \omega^{1/3}_0}{\left(v_F^2 q^2 + (\omega^{2/3}_q \omega^{1/3}_0)^2
\right)^{3/2}}.
\end{align}
With the change of variables  $\omega_q = \sqrt{v_F^3 z^3/\omega_0}$, the above expression becomes
\begin{align}
\nonumber \delta \Sigma_b (k,0) =  \frac{3k^2_F \e_k}{2N_F \left(2\pi\right)^2}\sqrt{\frac{1}{v_F \omega_0}} &\int_{0}^{\infty} dq dz  \frac{z^{3/2}q^2}{a^2 q^3 + \gamma \sqrt{\frac{v_F^3 z^3}{\omega_0}}} \\
& \times \frac{1}{\left(q^2 + z^2\right)^{3/2}}.
\end{align}
Converting this to polar coordinates, $q = r \cos \phi$ and $z = r \sin \phi$, we then have
\begin{align}
\delta \Sigma_b (k,0) =& \frac{3k^2_F \e_k}{2N_F \left(2\pi\right)^2}\sqrt{\frac{1}{v_F \omega_0}} \int_0^{\pi/2} d\phi \\
\nonumber &\times \int_{0}^{\infty} dr \frac{\cos^2 \phi \sin^{3/2} \phi}{a^2 r^{3/2} \cos^3 \phi  + \gamma \sqrt{\frac{v_F^3}{\omega_0}} \sin^{3/2} \phi}.
\end{align}
Integrating over $r$ and then over $\phi$, we find
\begin{align}
\delta \Sigma_b (k,0) = \e_k \frac{1}{\sqrt{3}} \left(\frac{k_F v_F}{a^4 \omega_0} \right)^{1/3} = \frac{2}{\pi} \e_k \approx 0.64 \e_k.
\end{align}
In earlier calculations of $\delta \Sigma_b (k,0) \propto \e_k$ in the quantum-critical regime near a $q=0$ instability ~\cite{rech2006quantum} it was assumed that the interaction is sufficiently long-ranged such that the instability develops when fermion-boson coupling $|U|$ is much smaller than the Fermi energy.  In this situation, the prefactor for $\e_k$ in $\delta \Sigma_b (k,0)$ is small in $(|U|/E_F)^{1/2}$, and $\delta \Sigma_b (k,0)$ can be neglected.  In our case, the interaction is short-ranged, critical $|U|$ is $1/N_F$, and $\delta \Sigma_b (k,0)/\e_k$ is a number of order one, which has to be kept when we determine $m^*/m$ and $Z$.

Using the above forms for the self-energy and solving the equation $G^{-1} (k, \omega_m) = i\omega_m -\e_k + \delta \Sigma (k, \omega_m) = (i\omega_m - \e^*_k)/Z$, we obtain
\begin{align}
\frac{m^*}{m} = \frac{1}{Z}\frac{1}{1-2/\pi}, ~~ \frac{1}{Z} = 1+\lambda = 1 + \frac{\xi k_F}{4a} \frac{Z m^*}{m}
\label{eq:s}
\end{align}
Solving this set we find that both $Z^{-1}$ and $m^*/m$ scale linearly with  $\xi$ at large $\lambda$, but with different prefactors. As a consequence, $Z m^*/m$ still tends to a constant value at $\xi = \infty$, but its value is different from one.
This changes $\lambda$ in $\delta \Sigma (k, \omega_m) = i \omega_m \lambda$ to  $\lambda  = \xi k_F/(1.45a)$
We also checked the corrections to the one-loop expression,  which come from series of vertex renormalizations.  The strongest ones at each loop order also scale as $\lambda \omega$ at large $k_F \xi$, but the prefactors do not form a particular series and very likely do not change qualitatively the one-loop result, i.e., $Z m^*/m$ remains of order one. 
Because the issue  here is only a number, which does not affect the functional form $\lambda \sim \xi k_F$,
 below we just set $Z m^*/m =1$ to simplify the calculations.

\subsection{$Z$  from the vertex function}

We now check which value of the quasiparticle  residue $Z$ we get if we use Fermi-liquid relations, Eq. (\ref{eq:1}), and use $V_{eff}$ for the vertex function.
The integral in the r.h.s of each of the two relations in (\ref{eq:1}) reduces to
\begin{widetext}
\begin{align}
 J= \int \frac{d^3 P}{(2\pi)^3} V_{eff} (K_F-P) \{G(P)^2\}_{\omega} = \frac{Z^2 k^2_F}{2 N_F}\int \frac{d^2 q}{(2\pi)^2}\frac{d\omega_q}{2\pi} \frac{1}{\xi^{-2} + (aq)^2 + \gamma
 \frac{|\omega_q|}{q}} \frac{1}{\left(i \omega_q - \vb{v}^*_F \cdot \vb q
\right)}\frac{1}{\left(i (\omega_q+\omega) - \vb{v}^*_F \cdot \vb q \right)} \Big \rvert_{\omega \rightarrow 0}
\end{align}
\end{widetext}
This is very similar to the expression for the self-energy. We verified that, like there,  the dominant contribution comes from the pole in the fermionic propagator. Evaluating the integral in the same way as we did for $\delta \Sigma_{f}$
 and using $Z m^*/m =1$,  we obtain
\begin{align}
J = \frac{Z^2 k^2_F}{2 v_F^* N_F} \int \frac{dq_y}{(2\pi)^2} \frac{1}{\xi^{-2} +(a q_y)^2} = Z \lambda.
\label{eq:9}
\end{align}
Substituting into (\ref{eq:1}), we obtain the self-consistent relation
\begin{align}
\label{rpa_Z} \frac{1}{Z} = 1 + Z \lambda
\end{align}
whose solution at $\lambda \gg 1$ is $Z \sim 1/\sqrt{\lambda}$.  This is different from $Z \sim 1/\lambda$, which we earlier extracted from the self-energy.

\section{Effective interaction and vertex function beyond RPA}

We now argue that at large $\lambda$, when $Z$ is small and $m^*/m$ is large,  one has to include Fermi liquid renormalizations in the calculation of the polarization $\Pi_{ph} (Q)$, which determines $V_{eff}$, and, most important,  include additional renormalizations of the vertex function, which make it different from $V_{eff}$. We will show below that using the renormalized vertex function instead of $V_{eff}$ restores the equivalence between the two calculations of the $Z$ factor.

 \subsection{Effective interaction}

We begin with $V_{eff}$.  We consider separately how Fermi-liquid renormalizations affect the Landau damping term, the $\xi^{-2}$ term, and the $(aq)^2$ term in (\ref{rpavertex}).

The Landau damping term comes exclusively from low-energy fermions with energies smaller than $\omega_q$, and the prefactor $\gamma$ can be computed using Eq. (\ref{eq:g}) for the Green's function.   The result is $\gamma = (Z m^*/m)^2 k^2_F/v_F$.
For $Z m^*/m =1$ the Landau damping term retains  the same value as for free fermions.

The $\xi^{-2}$ term also does not get renormalized. We recall that $\xi^{-2} = k^2_F(1-|U|\Pi (0))$, where $\Pi (0)$ is a static polarization bubble in the limit $q \to 0$. Because the self-energy in our case predominantly depends on frequency, $\Pi (0)$ can be computed by approximating $\int d^2k$ by $N_F \int d \e_k$ and integrating first over fermionic dispersion $\e_k$ over a range  $|\e_k| < W$, where $W$ is of order bandwidth:
\begin{align}
\nonumber \Pi(q \rightarrow 0) = - N_F \int_{-\infty}^{\infty} \frac{d \omega_m}{2\pi} \int_{-W}^{W} d
\e_k \frac{1}{\left(i (\omega_m + \Sigma (\omega_m))  -\e_k\right)^2}
\end{align}
The integral over $\e_k$ can be easily evaluated, and is determined by $|\e_k| \sim W$.  The corresponding $\omega_m$ are then also of order $W$.  At such high energies, the self-energy is irrelevant and $G(k, \omega)$ can be approximated by its free-fermion form.  Evaluating explicitly the frequency integral we then obtain $\Pi(0) = N_F$, like in RPA.  One can obtain the same result by integrating over $\omega_m$ first, but then one has to include contributions from both low-energy and high-energy fermions, and the two contributions partly cancel each other~\cite{maslov2010pomeranchuk}.

The $(aq)^2$ term has two contributions.  One comes from low-energy fermions, another from fermions with energies of order bandwidth~\cite{maslov2017gradient}.  The low-energy contribution gets  reduced by $m/m^*$ once we add Fermi liquid renormalizations~\cite{wolfle2011}. The high-energy contribution remains intact.  As a result, the prefactor $a$ gets reduced but remains finite, though for the cases considered in Ref.~\cite{maslov2017gradient}, this finite value is numerically quite small. Below we keep $a$ as a parameter, but comment on the case when $a$ is small.

The outcome of this analysis is that after Fermi liquid renormalizations, $V_{eff}$ retains the same functional form as in (\ref{rpavertex}).

\subsection{Vertex function}

\begin{figure}[t]
	\begin{center}
		\includegraphics[scale=0.14]{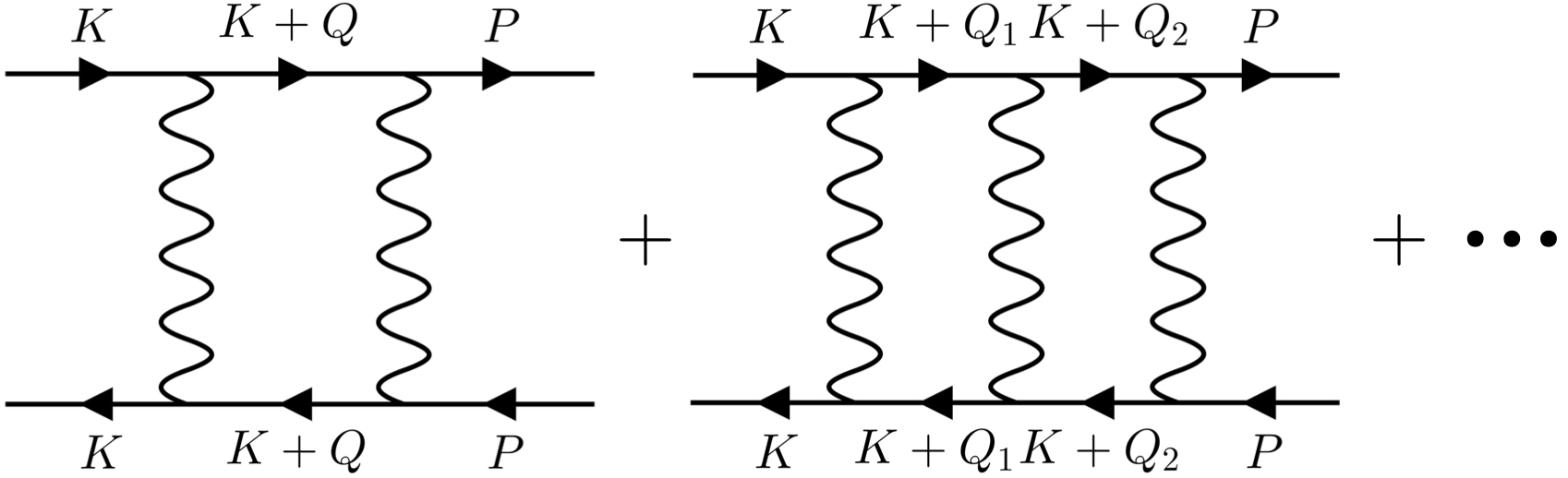}
		\caption{The sum over ladder diagrams using the effective RPA interaction. Note that these diagrams all
evaluate to zero when considering a static interaction, but due to the branch cut in the RPA vertex function, will
give finite results.}
		\label{ladder}
	\end{center}
\end{figure}

We next address the issue whether the equivalence between $\Gamma$ and $V_{eff}$ still holds near a QCP, when $\lambda$ gets large. 
We argue, following Refs.~\cite{maslov2010pomeranchuk,chubukov2014antiferromagnetic} that at large $\lambda$, there are additional relevant diagrams for the vertex function, which make it different from $V_{eff}$.  These additional diagrams contain a polarization bubble with zero momentum transfer (Fig. \ref{ladder}). In a weakly coupled Fermi liquid, these bubbles are convoluted with the static $U(q)$ and vanish after integration over frequency because they contain a double pole in either only upper or only lower frequency half-plane. Near a QCP, the product $G^2 (K+Q)$ is convoluted with the dynamical $V_{eff} (Q)$. 
The latter contains branch cuts in both half-planes of complex frequency, so the existence of a double pole in $G^2 (k+Q)$  in only one half-plane cannot be used to determine whether or not these contributions vanish. At this stage, it becomes essential that the vertex function is the $\omega$ limit of the dressed antisymmetrized interaction  $\Gamma(K, P, Q)$ (the limit of strictly zero momentum component of $Q$ and vanishingly small frequency component).
   
Let's consider the lowest-order diagram, with one insertion of $G^2 (K+Q)$ (left panel of Fig. \ref{ladder}). We label its contribution to $\Gamma$ as $\Gamma^{(2)}$ by the number of the interaction lines (in this the original $V_{eff}$ is $\Gamma^{(1)}$). Like we did before, we set external $K$ and $P$ on the Fermi surface, i.e., set  $\omega_k = \omega_p = 0$ and $|\vb k| = |\vb p|= k_F$.  
Using Fermi liquid form for $G(K+Q)$, we obtain
\begin{align}
\nonumber \Gamma^{(2)} =& \int \frac{d^2q}{(2\pi)^2} \frac{d \omega_q}{2\pi} G(K+Q)^2 V_{eff}(Q) V_{eff}(P-K-Q)\\
\nonumber = &\frac{k_F^4}{4N_F^2} \int \frac{d^2q}{(2\pi)^2} \frac{d\omega_q}{2\pi} \frac{Z^2}{(i \omega_q - \vb{v}_F^{*} \cdot \vb
q)^2}  \frac{1}{\xi^{-2} + a^2 q^2 + \gamma \frac{|\omega_q|}{q}} \\
& \times \frac{1}{\xi^{-2} + a^2 (\vb p -
\vb k - \vb q)^2 +\gamma \frac{|\omega_q|}{|\vb p - \vb k - \vb q|}}.
\label{eq:10}
\end{align}
The velocity $\vb{v}_F$ is directed along $\vb k$.  For definiteness, we set ${\vb k}$ along $x$. We assume and verify that relevant $|\vb k - \vb p|$ are of order $\xi^{-1}$, i.e., are much smaller than $k_F$, i.e., ${\bf k}_F$ and ${\bf p}_F$ nearly coincide. Then $\vb k -\vb p$ is directed almost along $y$ and its magnitude is approximately $k_F \theta$, where $\theta$ is the angle between $\vb k$ and $\vb p$, over which we will have to integrate later in Eq. (\ref{eq:1}).
The $G^2 (K+Q)$ has to be understood as the product of two Green's function with exactly the same momenta and infinitesimally small difference in frequencies, $\e >0$, i.e., $1/(i \omega_q - \vb{v}_F^{*} \cdot \vb q)^2  \to 1/((i \omega_q - \vb{v}_F^{*} \cdot \vb q) (i(\omega_q+ \e) - \vb{v}_F^{*} \cdot \vb q))$.
The integral in (\ref{eq:10}) is ultraviolet convergent and can be evaluated by integrating first either over frequency or over momentum.
Integrating over $q_x$ first, we find two contributions, $\Gamma^{(2)}_1$ and $\Gamma^{(2)}_2$. The first contribution comes from integrating over infinitesimally small $q_x$, when the poles in the two Green's functions are in different half-planes. This holds when $\omega_q$ is in the infinitesimally narrow range $-\epsilon < \omega_q <0$. This contribution then obviously comes from static $V_{eff}$, in which we can also set $q_x =0$.
In explicit form,
\begin{align}
\nonumber \Gamma^{(2)}_1 &= \frac{Z^2 k^4_F}{4 v_F^* N^2_F} \int_{-\infty}^{\infty} \frac{d q_y}{(2\pi)^2} \frac{1}{\xi^{-2} + a^2 q_y^2} \frac{1}{\xi^{-2} +
a^2 (k_F \theta + q_y)^2} \\
&= Z\lambda\frac{k^2_F}{2N_F} \frac{2}{4 \xi^{-2} + (a k_F\theta)^2}
\label{eq:f}
\end{align}
The second contribution, $\Gamma^{(2)}_2$ comes from the pole in the interaction, viewed as a function of $q_x$.
 For this contribution, one can set $\epsilon=0$.  For static interaction, the contribution from this pole cancels out $\Gamma^{(2)}_1$, as expected.
One can confirm this by taking $\Gamma^{(2)}_2$ with static $V_{eff}$ by integrating over $q_x$ first, in infinite limits, and closing the contour in the upper half-plane when $\omega_q<0$ and in the lower half-plane when $\omega_q>0$ in order to include only the contributions from the poles in the interaction. After performing subsequent elementary integrals over $\omega_q$ and then $q_y$, one obtains precisely the same result as for $\Gamma^{(2)}_{1}$ but with an overall minus sign.
However, in the presence of Landau damping term in $V_{eff}$, one can easily verify that $\Gamma^{(2)}_2$ is reduced
by $1/(k_F \xi)$ as characteristic  $q_y$ become of order $k_F$ rather than $\xi^{-1}$. Consequently, at large $\xi$, $\Gamma^{(2)}_2 \ll \Gamma^{(2)}_1$ and $\Gamma^{(2)} \approx \Gamma^{(2)}_1$, given by (\ref{eq:f}).

We also considered another second order diagram, one that takes the same form as $\Gamma^{(2)}$ but with the interactions crossed. We found that this diagram does not contain an overall factor of $\lambda$ compared to (\ref{eq:f}) and is therefore irrelevant. 

Extending this analysis to higher orders, we find that relevant diagrams form a ladder series. For an $mth$ order diagram we have
\begin{align}
\nonumber \Gamma^{(m)} =& \int \frac{d^2q_1}{(2\pi)^2} \frac{d \omega_1}{2\pi} \cdots \frac{d^2 q_{m-1}}{(2\pi)^2} \frac{d \omega_{m-1}}{2\pi}
G^2 (K+Q_1) \cdots \\ 
\nonumber \times & G^2 (K+Q_{m-1}) V_{eff}(Q_1) V_{eff}(Q_2-Q_1) \cdots \\
& V_{eff}(P-K-Q_{m-1}).
\end{align}
The largest contribution, of order $(Z\lambda)^{m-1}$, again comes from integration over infinitesimally small $q_y$ and $\omega_q$ in each cross-section, when the poles in the corresponding $G^2 (K+Q_i)$, with frequencies split by $\epsilon \to 0$, are in different half-planes. Integrating, we obtain
\begin{align}
\nonumber \Gamma^{(m)} =& \left(\frac{Z^2 m^*k_F }{8 \pi m}\right)^{m-1} \frac{k^2_F}{2N_F} \int dq_{1,y} \cdots \\
\nonumber &\times \int dq_{m-1,y} \frac{1}{\xi^{-2} +a^2 q_{1,y}^2} \\
\nonumber &\times \frac{1}{\xi^{-2} +a^2 (q_{2,y}-q_{1,y})^2} \cdots  \frac{1}{\xi^{-2}+a^2(k_F \theta
+ q_{m-1,y})^2}\\
\nonumber =& \left(\frac{Z^2\xi m^* k_F}{8 m a}\right)^{m-1} \frac{k^2_F}{2N_F} \frac{m}{m^2 \xi^{-2} + (a k_F \theta)^2} \\
=&\left(Z\lambda\right)^{m-1} \frac{k_F}{2N_F} \frac{m}{m^2 \xi^{-2} + (ak_F \theta)^2}
\end{align}
The total $\Gamma = \sum_{m=1}\Gamma^{(m)}$. For our purpose -- to extract $1/Z$ from $\Gamma$ using Eq.(\ref{eq:1}), the summation over $m$ can be simplified  because in (\ref{eq:1}) each $\Gamma^{(m)}$ is averaged over $\theta$. 
Because 
\begin{align}
\int d\theta \frac{m}{m^2 \xi^{-2} + (ak_F \theta)^2} = \int d \theta \frac{1}{\xi^{-2} + (a k_F \theta)^2},
\end{align}
$m/(m^2 \xi^{-2} + (ak_F \theta)^2)$ in each $\Gamma^{(m)}$ can be replaced by $1/(\xi^{-2} + (ak_F \theta)^2)$. The summation over $m$ is then elementary. 
Restoring spin indices, we obtain the full vertex function in the form
\begin{align}
\label{chargefull}
 \Gamma_{\alpha \beta, \gamma \delta} =  \Gamma^{RPA}_{\alpha \beta, \gamma \delta}~ Y
\end{align}
where $\Gamma^{RPA}_{\alpha \beta, \gamma \delta}$ is given by Eq. (\ref{ee_1}) and
\begin{align}
Y =  \sum_{m=1}^{\infty} \left(Z \lambda \right)^{m-1}
 =   \frac{1}{1- Z \lambda}
\label{eq:x5}
\end{align}
The relation is the same for charge and spin components of $\Gamma$.
Substituting into (\ref{eq:1}) we now obtain, instead of (\ref{rpa_Z}),
\begin{align}
\label{rpa_Z1} \frac{1}{Z} = 1 + \frac{Z \lambda}{1-Z \lambda} = \frac{1}{1-Z \lambda}
\end{align}
whose solution is $Z =1/(1+\lambda)$.
This is precisely the same result that we obtained from the self-energy.
We remind that $\lambda = (k_F \xi/4a) (Z m^*/m) \sim (k_F \xi/4a)$. Substituting into (\ref{eq:x5}) we find $Y = 1/(1- Z \lambda) =1/Z$.

Eqs. (\ref{chargefull})-(\ref{rpa_Z1}) are the central result of this paper. We caution, however, that this simple relation between the full $\Gamma (K,P)$ and $\Gamma^{RPA} (K,P)$ holds only for the vertex function at small momentum transfer between ${\vb k}$ and $\vb p$.  This is sufficient for the calculation of $1/Z$ using Eq. (\ref{eq:1}), but for a generic $|{\vb k}-\vb p| \sim k_F$, the computational procedure is more involved and the relation between $\Gamma$ and $\Gamma^{RPA}$ is more complex.

We also note that the ratio $m^*/m$ can be also obtained in a microscopic Fermi-liquid theory ~\cite{Chubukov2018}.  We don't present the calculations (they are similar to the ones for $1/Z$) and cite only the final result:
\begin{align}
\frac{m^*}{m} = 1 + \frac{Z\lambda}{1-Z \lambda}
\end{align}
Substituting $Z=1/(1+\lambda)$ we immediately find $m^*/m = 1 + \lambda = 1/Z$ as it should

\subsubsection{Additional term in the vertex function} 
There is one more addition due to dynamical nature of the effective interaction near a QCP.
We remind that the vertex function  $\Gamma^\omega (K,P; K,P)$ is the fully renormalized antisymmetrized interaction, i.e., the difference between fully renormalized interaction at zero momentum transfer and vanishingly small frequency transfer and  the fully renormalized interaction at finite momentum/frequency transfer $K-P$.
In the analysis above we obtained within RPA and then further renormalized the term in $\Gamma^\omega$ that depends on  the 2D  momentum transfer $K-P$  At the RPA level, there is no contribution to $\Gamma^\omega$ from vanishingly small 2D  momentum transfer because the corresponding polarization bubbles vanish in the $\omega$-limit.

\begin{figure}[t]
	\begin{center}
		\includegraphics[scale=0.035]{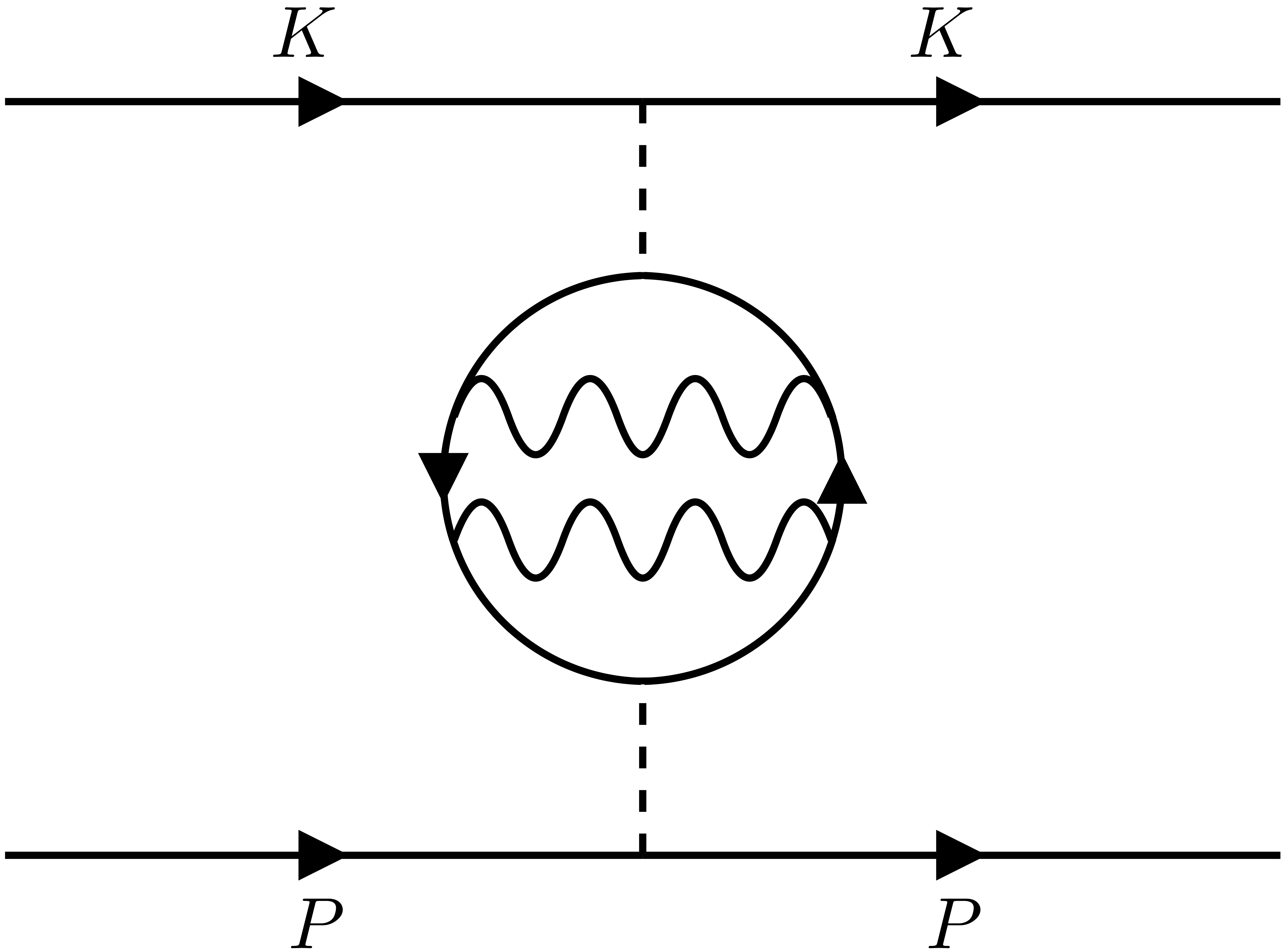}
		\caption{An example of a diagram that has zero momentum transfer and will contribute to only $F_0$, but none of the other Landau parameters or $Z$.}
		\label{zero_momentum}
	\end{center}
\end{figure}

The situation changes once we renormalize the bubble with zero momentum transfer by dynamical interaction $V_{eff}$. We show a representative of such diagrams in Fig. \ref{zero_momentum}.
Now the frequency integral does not vanish because of the branch cut in $V_{eff}$ in both half-planes of complex frequency.  Consequently, there is an additional contribution to charge component of $\Gamma^\omega (K,P)$ that does not depend on the 2D momentum transfer. We label this contribution as $C k^2_F/(2N_F)$. Combining with (\ref{chargefull}) we obtain the full $\Gamma^\omega$ near QCP in the form
\begin{widetext}
       \begin{align}
\label{chargefull_1}
\Gamma^\omega_{\alpha \beta, \gamma \delta} = &\frac{k^2_F}{4N_F} \left(\frac{1}{Z}
   \frac{1}{ \xi^{-2} + (a q)^2 + \gamma \frac{
 |\omega_q|}{ |\vb q|}} - C\right) \delta_{\alpha \gamma} \delta_{\beta \delta} + 
 \frac{k^2_F}{4N_F} \left(\frac{1}{Z}
   \frac{1}{ \xi^{-2} + (a q)^2 + \gamma \frac{
 |\omega_q|}{ |\vb q|}}\right) \vec{\sigma}_{\alpha \gamma} \vec{\sigma}_{\beta \delta}.
\end{align}
\end{widetext}
The constant term $C$ does not affect Eq. (\ref{eq:1})  because the integral in the r.h.s of (\ref{eq:1}) vanishes once we substitute a constant for $\Gamma$. It also does not affect the Fermi liquid relation between $\Gamma$ and  $m^*/m$. 
However, we show in the next section that this term is essential for proper calculation of the Landau parameter $F_0$ in a critical Fermi liquid

\section{Critical Fermi liquid}

We now describe the implementation of our expressions for $V_{eff}$ and $\Gamma$ for a Fermi-liquid theory near a QCP when $\lambda$ is large, and the dynamics of $V_{eff}$ plays a crucial role in self-consistent calculations of $Z$ and $m^*/m$. Following~\cite{maslov2010pomeranchuk}, we call this a critical Fermi liquid in order to distinguish it from a conventional Fermi liquid, for which $V_{eff}$ can be approximated by its static form. In our notation, conventional Fermi liquid regime holds when $\lambda <1$ and critical Fermi liquid regime holds for $\lambda >1$.

The two issues we discuss here are (i) the expressions for Landau parameters $F_n$ and $G_n$ in charge and spin channels with angular momentum $n$ and (ii) the forms of static charge and spin susceptibilities $\chi_{c,n}$ and $\chi_{s,n}$.

The Landau parameters $F_n$ and $G_n$ are partial components of the Landau function $f_{\alpha \beta, \gamma \delta} (|\vb k- \vb p|)$, in which both momenta are on the Fermi surface. The Landau function $f$  is  related to $\Gamma^\omega$ as
\begin{align}
f_{\alpha \beta, \gamma \delta} (|\vb k_F - \vb p_F|) = 2 N_F Z^2 \frac{m^*}{m} \Gamma^\omega (K_F,P_F)
\label{eq:k}
\end{align}
with $K_F=(\vb k_F,0)$. The partial components $F_n$ and $G_n$ are obtained by integrating charge and spin components of $f$ with $\cos(n \theta)$, where $\theta$ is the angle between Fermi surface momenta  ${\bf k}$ and ${\bf p}$.

For partial components with $n>0$, the constant term in $\Gamma^\omega$ in (\ref{chargefull_1}) is irrelevant as the $\int d \theta \cos{n \theta}$  vanishes. Keeping  the momentum-dependent terms in (\ref{chargefull_1})
 and using $Z m^*/m =1$,
we obtain for $n>0$
 \begin{align}
F_n = G_n =  \frac{k^2_F}{4\pi} \int d \theta \frac{\cos{n\theta}}{\xi^{-2} + 2(ak_F)^2 (1-\cos{\theta})}
\label{eq:x}
\end{align}
For not too large $n < n_c \sim \lambda a^2$,
\begin{align}
F_n = G_n \approx  \frac{k_F \xi}{4a} = \lambda
\label{eq:x1}
\end{align}
We see that Landau parameters scale as $\lambda$, i.e., diverge upon system approach to a QCP.
 A similar result has been obtained in Ref.~\cite{maslov2010pomeranchuk} for a critical Fermi liquid near a nematic QCP. 

The corresponding susceptibilities are ~\cite{leggett1965, lifshitz1980statistical}
\begin{align}
\chi_{n,c}  = \chi^{(0)}_{n,c} \frac{\frac{m^*}{m}}{1+ F_{n}} + 
\chi^{inc}_{c,n};  \chi_{n,s}  = \chi^{(0)}_{n,s} \frac{\frac{m^*}{m}}{1+ G_{n}} + \chi^{inc}_{s,n},
\label{eq:x2}
\end{align}
where $\chi^{(0)}_{n,c}$ and  $\chi^{(0)}_{n,s}$ are partial susceptibilities of free fermions and $\chi^{inc}$ are the component of susceptibility that comes from high-energy fermions, outside the applicability range of fermi-liquid theory.
These terms do not contain divergent $m^*/m$ in the prefactor and are  therefore irrelevant to our purposes.
 
We see from (\ref{eq:x2}) that $1+F_n = 1+G_n = 1+\lambda$ cancel out singular mass renormalization $m^*/m = 1 +\lambda$. As a result, all non-s-wave susceptibilities remain finite at an $s-$wave QCP.  This result is fully expected on physical grounds. We emphasize that the presence of the extra factor $1/Z$ between $\Gamma$ and $V_{eff}$ in (\ref{chargefull}) plays the crucial role here. Without such factor, we would not have found a cancellation between $m^*/m$ and $1 + F_n$ or $1+G_n$.

The same analysis holds for $s-$wave spin component. We have $G_0 \approx \lambda$, such that  $\chi_{0,s}$ remains approximately equal to susceptibility of free fermions.

For the $n=0$ charge component we obtain
\begin{align}
F_0 \approx \lambda - \frac{C}{2} \frac{k^2_F}{1 + \lambda}
\label{eq:x3}
\end{align}
At an $s-$wave charge QCP we must have $F_0 =-1$. This requirement is met if
$C  = 2(1+\lambda)^2/k^2_F + O(1/(ak_F)^2)$,
which we assume to hold.
Substituting into (\ref{eq:x3}) we obtain
\begin{equation}
1 + F_0 \sim \frac{1}{a^2 (1+ \lambda)}  \sim \frac{1}{a k_F \xi}.
\label{sss}
\end{equation}
Substituting this into the expression for the charge susceptibility and using  $m^*/m = 1 + \lambda$, we  then obtain
\begin{align}
\chi_{0,c} = \chi^{(0)} \frac{\frac{m^*}{m}}{1+ F_{0}} \sim \chi^{(0)}  \left((1+\lambda) a\right)^2 \sim
\chi^{(0)} (k_F \xi)^2
\label{eq:x4}
\end{align}
where $\chi^{(0)}$ is the susceptibility of free fermions.  We see that the $s-$wave charge susceptibility diverges as $\xi^2$ near a QCP.   The divergence is the same as we would obtain within RPA, by extracting charge susceptibility from the effective interaction $V_{eff} (Q=0) \propto \xi^2$.  We see however that the divergence is now split between $m^*/m \sim \xi$ and $1/(1+F_0) \sim \xi$. In other words, as long as RPA description is valid (i.e., as long as $\lambda \leq 1$),  $m^*/m \approx 1$ and $(1+F_0) \sim \xi^{-2}$.  Closer to QCP, when $\lambda$ increases, $m^*/m$ becomes large and simultaneously the slope of $(1+F_0)$ changes from $\xi^{-2}$ to $\xi^{-1}$. 

This behavior becomes more exotic in the limit when the prefactor $a$ in the RPA expression for $V_{eff}$ by itself scales as $(k_F \xi)^{-1}$, which, as we said,  happens when it predominantly comes from low-energy fermions. In this situation, $m^*/m \sim \lambda \sim \xi^2$, while $(1 + F_0) \sim 1/(a k_F \xi)$  saturates at a constant value instead of vanishing.  The charge susceptibility still diverges at a QCP as $\xi^2$, but the divergence now comes exclusively from the effective mass. This limit, however, is somewhat artificial as at $a \to 0$ susceptibilities in all channels diverge at a QCP, i.e., a critical point becomes multidimensional.

\section{Conclusions}

In this communication, we considered a system close to a $q=0$ charge QCP.
We have shown here that near a QCP, when the correlation length $\xi$ for charge fluctuations is large compared to $1/k_F$,  the system enters into a critical Fermi liquid regime. In this regime, excitations are still coherent at the smallest frequencies and the quasiparticle $Z$-factor tends to a finite value at $\omega \to 0$.  But the vertex function -- the one that determines $Z$, the mass renormalization, and the thermodynamic properties at the lowest energies, becomes different from the ones in a conventional Fermi liquid.  Specifically, the ``forbidden" contributions to the vertex function, which vanish in conventional Fermi-liquid theory, now become non-zero because the dynamics associated with Landau damping now play a crucial role.  Near a QCP, summing up a ladder series of these formally forbidden contributions increases the vertex function by $1/Z \propto (\xi k_F)$.
We demonstrated that the $Z$ factor, obtained self-consistently from charge and spin Ward identities within a microscopic Fermi-liquid theory, is equivalent to $Z$ extracted from the quasiparticle self-energy.  The extra ladder diagrams are crucial for this equivalence. We also argued that the charge component of the full vertex function $\Gamma^\omega$ has an additional constant term with a prefactor that scales as $\xi^2$. This additional term does not affect the Fermi liquid relations between $1/Z$ (and $m^*/m$) and the vertex function, but must be kept to satisfy the condition that the Landau parameter $F_0$ must approach $-1$ at the QCP. All other Landau parameters $F_n$ and $G_n$ diverge near a QCP in the same way as $m^*/m$.
The cancellation of divergencies between $m^*/m$ and $1 + F_n$ for charge susceptibilities and  $1+ G_n$ for spin susceptibilities ensures that these susceptibilities remain finite at an $s-$wave charge QCP, in line with common sense reasoning.

Our results are consistent with prior works  in which the full $\Gamma$ was obtained in a critical FL near a $d-$wave nematic order~\cite{chubukov2009ferromagnetic,maslov2010pomeranchuk} and $(\pi,\pi)$ antiferromagnetic order~\cite{chubukov2014antiferromagnetic}. Compared to these previous cases, the s-wave charge QCP allows for explicit  evaluation of the vertex function and yields results that illustrate the problem with more clarity. The analytical calculations presented here are controlled, and the solution explicitly resolves the apparent discrepancy found at the level of RPA  between the quasiparticle residue $Z$ extracted from the fermionic self-energy and from the vertex function.

\section{Acknowledgments}
We thank D.~Maslov for helpful discussions. 
The work of was supported  by U.S. Department of Energy, Office of Science, Basic Energy Sciences, under Award No. DE-SC0014402. A.C.  acknowledges the hospitality of the Kavli Institute for Theoretical Physics (KITP), Santa Barbara, supported by the National Science Foundation under Grants No.~NSF PHY-1748958 and PHY-2309135.

\bibliography{charge_bib}

\end{document}